\documentclass[twocolumn,showpacs]{revtex4}
\usepackage[xdvi]{graphicx}%
\usepackage{dcolumn}
\usepackage{amsmath}
\makeatletter
\def\btt#1{\texttt{\@backslashchar#1}}%
\DeclareRobustCommand\bblash{\btt{\@backslashchar}}%
\makeatother

\begin{document}
\title{Steady state thermodynamics for heat conduction}
\author{Shin-ichi Sasa and Hal Tasaki}
\affiliation
{Department of Pure and Applied Sciences,
University of Tokyo, Komaba, Tokyo 153-8902, Japan,
Department of Physics, Gakushuin University,
Mejiro, Toshima-ku, Tokyo 171-8588, Japan}

\date{\today}

\begin{abstract}
Following the proposal of
steady state thermodynamics
(SST) by
Oono and Paniconi,
we develop a phenomenological theory
for steady nonequilibrium states in systems
with heat conduction.
We find that there is essentially a unique
consistent thermodynamics, and
make concrete predictions, i.e, the existence
of a new osmotic pressure and a shift in
the coexistence temperature.
These predictions allow one to test for
the quantitative validity of SST
by comparing them with experiments.
\end{abstract}

\maketitle

Construction of a statistical mechanics that apply to
nonequilibrium states
has been a challenging open problem in theoretical physics
\cite{RP}.
But so far it is not known how the desired
probability measures for nonequilibrium states look like,
or even whether the measures can be written in
compact forms
\cite{fn:measure}.
Recalling the history that the conventional thermodynamics
was an essential guide when Boltzmann, Gibbs, and others
constructed equilibrium statistical mechanics,
it may be a good idea to start from the level of
thermodynamics.

The standard theory of nonequilibrium
thermodynamics \cite{LE}
is based on local equilibrium hypothesis, which roughly
asserts that each small part of a nonequilibrium state can be regarded
as a copy of a suitable equilibrium state.
But such a description seems
insufficient for general
nonequilibrium states. Consider, for example, a system
with a steady heat flow. It is true that the quantities like the temperature
and the density become essentially constant within a sufficiently small
portion of the system. But no matter how small the portion is, there always
exists a heat flux passing through it and hence the local state  is
{\em not}\/ isotropic. This suggests that the local state cannot be identical
to an equilibrium state (which is isotropic), but should be described rather
as a {\em local steady state}\/.

Among existing attempts in
nonequilibrium thermodynamics
to go beyond local equilibrium treatments \cite{NLE},
the steady state thermodynamics (SST)
proposed by Oono and Paniconi \cite{OP}
seems to be most sophisticated and
promising.
The basic strategy of \cite{OP},
in our own interpretation,
is to
i)~look for a thermodynamics
which describes
a steady state as a whole,
ii)~clarify operational procedures for
determining thermodynamic quantities,
and
iii)~respect the general mathematical
structure of thermodynamics.
In the present Letter, we apply
this strategy
to a concrete problem of heat
conduction (in a fluid),
and show that theoretical consistency
leads one to an essentially
unique thermodynamics.
We then make some concrete predictions
which may be confirmed quantitatively
in experiments.
Extensions to other systems
and discussions of related microscopic results
will appear in
\cite{ST2}.

\begin{figure}
\begin{center}
\includegraphics[width=6cm]{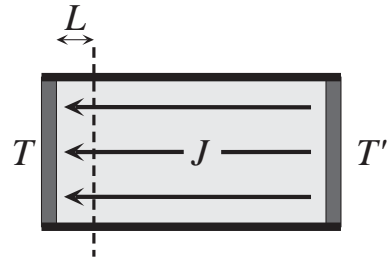}
\end{center}
\caption{
A typical system treated in SST.
The left and the right walls have temperatures
\( T \) and \( T' \), respectively,
and there is a steady heat flux \( J \).
We get a local steady state
by restricting our attention to the thin
portion within a distance \( L \) from
the left wall (as denoted by the dashed line)
}
\label{f:sys}
\end{figure}

\paragraph*{Local steady state:}
We consider a macroscopic
system of a single
substance in a closed
cylindrical container
with the cross section area \( A \).
The left and right walls of the container
have very efficient heat
conduction, and are kept at constant
temperatures \( T \) and \( T' \),
respectively, with the aid of
external heat baths.
See Fig.~\ref{f:sys}.
The side walls of the container
are perfectly adiabatic.

If the system is kept in this
setting for a sufficiently long
time, it is expected to reach
a unique steady state with
a constant heat current
without any
macroscopically
observable changes.
(We assume that convection
does not take place.)
By \( J \) we denote the total
energy that flow into the system from
the right wall within a unit time.

We now restrict our attention to a thin
portion of the system within a small distance
\( L \) from the left wall
as in Fig.~\ref{f:sys}.
Here the length \( L \) is taken so that the
temperature in the thin portion becomes
essentially constant.
We assume (as in local equilibrium
approaches) this is realized with \( L \)
which is much larger than any microscopic
scales.
The state in this thin portion is the
{\em local steady state}\/
that we study \cite{note:geometry}.
Our first crucial assumption is that
the local steady state can be fully
specified by four macroscopic
parameters as \( (T,J;V,N) \),
where \( V=AL \) is the volume
and \( N \) is the amount of substance
in the thin portion.
We stress that the heat flux
\( J \) need not be small.

As in the conventional thermodynamics,
it is essential to consider
decomposition/re\-combination
and scaling
of local steady states.
In doing so, we shall always fix the cross section
\( A \) and vary only the length \( L \)
(within the range the system remains thin).
Consider splitting the system of length
\( L \) into those with lengths
\( L_{1} \) and \( L_{2} \)
(with \( L=L_{1}+L_{2} \))
by a plane parallel to the
left and right walls \cite{note:plane}.
Correspondingly, we assume that the local steady state
\( (T,J;V,N) \) can be decomposed into two
local steady states
\( (T,J;V_{1},N_{1}) \)
and
\(  (T,J;V_{2},N_{2}) \), and
the two states
can be recombined
back into
\( (T,J;V,N) \).
Here \( V_{i}=AL_{i} \),
and
\( N_{1}+N_{2}=N \).
Similarly, for \( \lambda>0 \),
we assume that
one can scale the length \( L \)
to \( \lambda L \) to
get a scaled copy
\( (T,J;\lambda V,\lambda N) \)
of the state
\( (T,J;V,N) \).
These observations imply
that
the heat flux \( J \)
behaves in a similar way as \( T \), and
hence
should be
regarded as an
``intensive'' variable.
This identification is essential in our
theory.

\paragraph*{SST free energy:}
Our second essential assumption
is the existence of the
SST free energy \( F(T,J;V,N) \),
which is a thermodynamic potential
describing the response of the
local steady state when the extensive
variables \( V \) and \( N \) are
varied.
We assume that there is
a function  \( F(T,J;V,N) \)
which is concave in intensive
variables \( T \), \( J \),
and convex in extensive
variables \( V \), \( N \).
It should have {\em additivity}\/
 \( F(T,J;V,N)
 =
 F(T,J;V_{1},N_{1})+F(T,J;V_{2},N_{2}) \)
for any decomposition as above,
{\em extensivity}\/
\( F(T,J;\lambda V,\lambda N)
=\lambda\,F(T,J;V,N) \)
for \( \lambda>0 \),
and {\em symmetry}\/
\( F(T,J;V,N)=F(T,-J;V,N) \)
to reflect the obvious left-right
symmetry of thin systems.
Most importantly, we require
that the derivatives of
\( F(T,J;V,N) \) in \( V \) and \( N \) have
operational meanings exactly parallel to
those in the conventional
thermodynamics.

\begin{figure}
\begin{center}
\includegraphics[width=8cm]{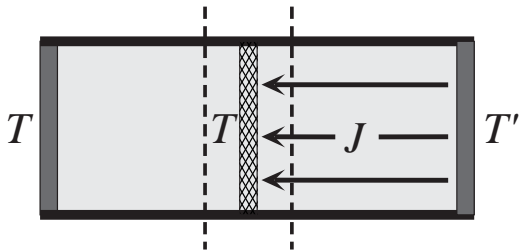}
\end{center}
\caption{
The porous wall in the middle of the container
separates an equilibrium state
and a steady nonequilibrium state.
If we restrict our attention to the vicinity
of the porous wall
(denoted by dashed lines),
we get a situation where
a local steady state
and an
equilibrium state
are in balance with each
other.
We use this setting to measure the chemical potential
of the steady state.
We will also see that
there appears a force (of nonequilibrium origin)
that pushes the porous wall towards the equilibrium
region irrespectively to the sign of \( J\ne0 \).
This is an example of
``flux induced osmosis.''
}
\label{f:FIO}
\end{figure}

More precisely, we first require
\( \partial F(T,J;V,N)
/\partial V = -p(T,J;V,N) \),
where the pressure
\( p(T,J;V,N) \) is simply
determined by measuring the force
that the system exerts on
the left (or the right) wall \cite{fn:walls}.
We then require
\( \partial F(T,J;V,N)
/\partial N = \mu(T,J;V,N) \),
where the existence and measurability of
the chemical potential
\( \mu(T,J;V,N) \)
are assumed.
A device \cite{fn:mu} for measuring
\( \mu(T,J;V,N) \)
is depicted in Fig.~\ref{f:FIO}.
Here the left and the right walls of the container
are kept at constant temperatures
\( T \) and \( T' \), respectively.
In the middle of the container, there
is another wall made of a porous medium
which is kept at a constant temperature
\( T \) with the aid of an external heat bath.
We assume that the substance can move
across the porous wall.
Suppose that
the whole system has reached a steady state.
Then the left half of the system has a constant
temperature, and is in a normal
equilibrium state (provided that the
middle walls is in a very efficient contact
with a bath).
The right half is in a nonequilibrium
steady state
(which is not necessarily local)
with a constant heat flux.
In order to examine the balance between
the two parts, we restrict our attention to
a thin part of the system within a fixed
distance from the porous wall
as  denoted by dashed lines
in Fig.~\ref{f:FIO}.
If the distance is small enough,
the part of the steady state
can be regarded as local.
Then we get a situation where
a local steady state
\( (T,J;V,N) \)
and an
equilibrium state
\( (T,0;V',N') \)
are in balance with each
other with respect to the exchange
of the substance.
Then it is natural to define
\( \mu(T,J;V,N) \) to be
equal to \( \mu(T,0;V',N') \),
where the latter
can be determined
within the conventional
equilibrium thermodynamics.
The two derivatives and the extensivity
determine \( F(T,J;V,N) \)
without any ambiguities \cite{fn:2nd},
as is obvious from the Euler equation
\begin{equation}
    F(T,J;V,N)=-V\,p(T,J;V,N)+N\,\mu(T,J;V,N),
    \label{eq:Euler}
\end{equation}
which is derived by using the extensivity as usual.

Let us define (extensive) SST entropy as
\begin{equation}
    S(T,J;V,N)=
    -\frac{\partial F(T,J;V,N)}
    {\partial T},
    \label{eq:S}
\end{equation}
and a new ``extensive'' quantity
\begin{equation}
    \Psi(T,J;V,N)=
    -\frac{\partial F(T,J;V,N)}
    {\partial J},
    \label{eq:Psi}
\end{equation}
which we shall call
{\em nonequilibrium order parameter}\/.
{}From the symmetry and the concavity of \( F \),
one finds that
\( \Psi(T,J;V,N)=-\Psi(T,-J;V,N)\ge0 \)
for \( J\ge0 \).
We expect to have
\( \Psi(T,J;V,N)>0 \) for \( J>0 \)
in generic systems.
Note that
both \( S \) and \( \Psi \)
can be
measured operationally since
\( F \) can be.
As in the conventional thermodynamics, one can derive
various identities between thermodynamic
quantities.

The existence of the SST free energy
with the desired properties is
nothing more than an optimistic
assumption.
It is possible in principle that such a theoretical
framework as SST simply does not exist in
Nature.
We therefore predict two concrete phenomena
--- the existence of a new
osmotic pressure called
flux-induced osmosis (FIO) and a
shift of coexistence temperature ---
and present some exact relations,
which enable one to test for the
{\em quantitative}\/ validity of
SST through experiments.

\paragraph*{Flux-induced osmosis:}
Let us consider the situation in
Fig.~\ref{f:FIO},
and examine the behavior of the pressure
\( p_{\rm ss}=p(T,J;V,N) \)
of the steady state.
We fix the temperature
\( T \) and the pressure
\( p_{\rm eq}=p(T,0;V',N') \)
(by suitably varying the volume
\( V' \))
of the
equilibrium state,
and vary only the flux \( J \).
The chemical potential of the equilibrium
state is thus constant, and so is the
chemical potential \( \mu(T,J;V,N) \)
of the local steady state
(by definition).
We differentiate the Euler equation
(\ref{eq:Euler})
by \( J \) with \( T \) and \( p_{\rm eq} \)
fixed.
Keeping in mind that \( V \) and \( N \)
may depend on \( J \),
we find
\begin{equation}
    \frac{\partial p_{\rm ss}(T,p_{\rm eq},J)}
    {\partial J}
    =
    \frac{\Psi(T,J;V,N)}{V}.
    \label{eq:FIO}
\end{equation}
Noting that
\( p_{\rm ss}(T,p_{\rm eq},0)=p_{\rm eq} \)
and recalling the sign of \( \Psi \), this
implies
\( p_{\rm ss}\ge p_{\rm eq} \) in general.
If \( \Psi \) is nonvanishing
(as we expect)
then one has
\( p_{\rm ss} > p_{\rm eq} \)
for \( J\ne0 \).
Recalling that the pressures are defined
from mechanical forces exerted on the walls,
this implies that {\em the porous wall is actually
pushed towards the equilibrium region
irrespectively to the sign of
the heat flux}\/ \( J \).
We stress that this force, which is absent in
equilibrium or local-equilibrium treatments,
is of purely nonequilibrium origin \cite{fn:aniso}.
This is an example of a general
phenomenon that we call
{\em flux-induced osmosis}\/ (FIO) \cite{ST2}.

In the same situation, one can also
derive  \cite{ST2} a nontrivial identity
\begin{equation}
\frac{\partial p_{\rm ss}(T,p_{\rm eq},J)}{\partial p_{\rm eq}}
=\frac{v_{\rm eq}}{v_{\rm ss}},
\end{equation}
where \( v_{\rm eq}=V'/N' \)
and \( v_{\rm ss}=V/N \).
Since the identity
involves only directly measurable
quantities, it
may be useful in quantitative
tests of SST
in (real and numerical)
experiments.

\paragraph*{Shift of coexistence temperature:}
Consider again the original setting in
Fig.~\ref{f:sys}, and assume that the pressure
\( p \) is kept constant.
We further assume that a phase coexistence takes
place in the system,
i.e., the lower temperature region
of the container is
occupied by one phase (e.g. liquid)
while the higher temperature region
by another (e.g. gas).
We then ask what is the temperature
\( T_{\rm c}(p,J) \) at the boundary
between the two phases.
Within local equilibrium treatments,
one simply concludes that \( T_{\rm c}(p,J) \)
is the same as its equilibrium value
\( T_{\rm c}(p,0) \).
In SST, however, we find
(from an analysis similar to that for FIO)
that
\begin{equation}
    \frac{\partial T_{\rm c}(p,J)}
    {\partial J}
    =
    -\frac{\psi_{\rm high}-\psi_{\rm low}}
    {s_{\rm high}-s_{\rm low}},
    \label{eq:shift}
\end{equation}
were \( \psi_{\rm low} \) and
\( \psi_{\rm high} \)
are the molar \( \Psi \)
of the two phases at the
coexistence point, and
\( s_{\rm low} \) and \( s_{\rm high} \)
are the corresponding molar
SST entropies.
The identity (\ref{eq:shift}) means
that {\em in general
\( T_{\rm c}(p,J) \) shifts from
its equilibrium value}\/.
In case
\( T_{\rm c}(p,J)>T_{\rm c}(p,0) \)
\cite{fn:shift},
one has a remarkable phenomenon of ``heat flux induced
freezing'', i.e., one observes a solid phase in a system
with one wall having a temperature slightly
higher than the melting point and the other
wall having a much higher temperature.

\paragraph*{Choice of nonequilibrium variable:}
Finally
let us make an important remark about the choice of
nonequilibrium thermodynamic variable.
A local steady state
may also be specified as
\( (T;V,N,\tau) \),
where  \( \tau \)
is the (small) temperature
difference between the left and right ends of
the system.
Since \( \tau \) is proportional to the length
\( L \) in a thin uniform state, it should be
regarded as an ``extensive'' quantity.
Then, in any mathematically ``healthy''
thermodynamics, the corresponding free energy
\( \tilde{F}(T;V,N,\tau) \)
(if exists) should be
convex in \( V \), \( N \), and \( \tau \).

Following the standard argument,
the convexity and the extensivity implies
a variational principle
\begin{eqnarray}
    &&
    \tilde{F}(T;V_{1},N_{1},\tau^*)
    +
    \tilde{F}(T;V_{2},N_{2},\tau-\tau^*)
    \nonumber\\
    &&
    =
    \min_{\tau'}
    \{
    \tilde{F}(T;V_{1},N_{1},\tau')
    +
    \tilde{F}(T;V_{2},N_{2},\tau-\tau')
    \},
    \label{eq:tauvar}
\end{eqnarray}
for fixed \( V_{1} \), \( V_{2} \), \( N_{1} \),
\( N_{2} \), and \( \tau \).
This relation determines the temperature
\( T+\tau^* \)
at the boundary of the two parts of the system
(with volumes \( V_{1} \) and \( V_{2} \), respectively)
when the total temperature difference \( \tau  \)
is fixed.
In a differential form, this condition becomes
\( \nu(T;V_{1},N_{1},\tau^*)=
\nu(T;V_{2},N_{2},\tau-\tau^*) \),
where
\( \nu(T;V,N,\tau)=
\partial\tilde{F}(T;V,N,\tau)
/\partial\tau \).

On the other hand, we already know from the energy
conservation law that \( \tau^* \) can
be determined by the condition
\( J(T;V_{1},N_{1},\tau^*)=
J(T;V_{2},N_{2},\tau-\tau^*) \),
where we expressed the heat flux \( J \)
as a function of \( (T;V,N,\tau) \).
Since the conditions written in \( \nu \)
and in \( J \) must be equivalent,
we conclude that there is a function
\( f \) such that
\( J(T;V,N,\tau)=f(\nu(T;V,N,\tau)) \)
for any \( (T;V,N,\tau) \).
But, noting that \( J \) has a dimension of
energy divided by time while
\( \nu \) is dimensionless,
we find that such a universal
function \( f \) simply
does not exist.
This  observation implies that {\em one can never get
a consistent thermodynamics by using the temperature
difference \( \tau \) as a nonequilibrium variable}\/.
In fact we believe that the representation in terms
of \( (T,J;V,N) \),
where the nonequilibrium variable \( J \)
directly reflects the energy conservation law,
provides
an essentially unique consistent
thermodynamics for a system with steady heat flux
\cite{fn:another}.

\paragraph*{Discussions:}
By following the
philosophy of \cite{OP},
we are led to an essentially unique
thermodynamics
(SST) for heat conduction.
The uniqueness suggests that, if a consistent
thermodynamics for steady heat conduction
exists at all, then it should be equivalent to
what we have described here.
The
resulting theory led us to
novel predictions
which allow one to
test for the quantitative validity of
SST through experiments, other theoretical
results \cite{fn:theo}, and numerical calculations
\cite{fn:num}.

Let us stress that SST is by no means
incompatible with
the long range (spatial) correlation
generically found in nonequilibrium steady
states \cite{LRC}.
The long range correlation comes
from the global description based on fluctuating
hydrodynamics, while local aspects
are conventionally described as local steady states.
The aim of SST is to provide a more sophisticated
alternative of the latter, i.e., to give
a unified description of nonequilibrium
modifications of the (local) equations of states.
Since there is a big separation in the length scales,
the existence of the long range correlation
does not conflict with the validity of
SST.

Perhaps the most crucial point about
the potential significance of
SST is whether
it becomes a useful guide in the (future)
construction of statistical mechanics
for steady nonequilibrium states.
The fact that we have arrived at an
essentially unique theory
is rather encouraging.
We hope that, by trying to construct a statistical
theory that is consistent with the (unique)
nonequilibrium thermodynamics,
we are naturally led to a meaningful and
correct statistical mechanics
for steady nonequilibrium states.

It is a pleasure to thank Yoshi Oono for
introducing us to the present subject and
for various indispensable advices and suggestions.
The work of S. S.
was supported by grants from the Ministry of Education, Science,
Sports and Culture of Japan, Nos. 12834005 and  11CE2006.


\begin{thebibliography}{10}
\bibitem{note:geometry}
We have here restricted ourselves to the simplest
geometry.
We plan to treat a more general class of
steady states by ``connecting'' together
local steady states.
\bibitem{note:plane}
One may regard the plane as being purely fictitious, or
as a highly conducting wall kept at a suitable
temperature with the aid of a heat bath.
\bibitem{fn:walls}
Note that the forces exerted on the
left and the right walls are exactly the same
(even when the system is not ``thin'').
Otherwise the system gets a finite acceleration
and cannot be steady.
\bibitem{fn:mu}
This is not the unique
way of determining \( \mu \).
Another natural method is to put a system
into a direct contact with a particle bath.
\bibitem{fn:shift}
The concavity of the Gibbs free energy
implies
\( s_{\rm low}< s_{\rm high} \) in general.
We further expect to have
\( \psi_{\rm low}>\psi_{\rm high}  \)
in many systems.
\bibitem{fn:another}
In fact one can think about another thermodynamics
in which one fixes the length \( L \) and
vary only the cross section \( A \).
Then one should regard the temperature difference
\( \tau \) as ``intensive'' and
its conjugate \( \nu \) as ``extensive.''
But this theory may not be as robust as the one we propose here
\cite{ST2}.
\end{thebibliography}
\end{document}